\documentclass[12pt]{article}

\usepackage[english]{babel}
\usepackage[utf8x]{inputenc}
\usepackage{amsmath}
\usepackage{graphicx}
\usepackage[colorinlistoftodos]{todonotes}
\usepackage{subfigure}

\title{Scattering amplitudes in Lifshitz spacetime}
\author{Tom\'as Andrade\footnote{tomas.andrade@durham.ac.uk}, Yang Lei\footnote{yang.lei@durham.ac.uk}, and Simon F. Ross\footnote{s.f.ross@durham.ac.uk} \\ \bigskip \\ Centre for Particle Theory, Department of Mathematical Sciences \\ Durham University \\ South Road, Durham DH1 3LE}

\begin{document}
\maketitle

\begin{abstract}
We consider the calculation of scattering amplitudes in field theories dual to Lifshitz spacetimes. These amplitudes provide an interesting probe of the IR structure of the field theory; our aim is to use them to explore the observable consequences of the singularity in the spacetime. We assume the amplitudes can be related by T-duality to a Wilson loop, as in the AdS case, and determine the bulk minimal surfaces for the simplest cusp Wilson loop. We use this to determine the leading IR singularity in the amplitude. We find there is a stronger IR singularity for $z >1$ than for $z=1$, with a coefficient which vanishes as $z \to 1$.
\end{abstract}

\section{Introduction}

The application of holography to non-relativistic theories represents an interesting extension of the usual holographic dictionary. The simplest example of this type is the Lifshitz spacetime \cite{Kachru:2008yh}. The geometry is
\begin{equation} \label{Lifspace}
ds^2=-\frac{dt^2}{r^{2z}}+\frac{dr^2+d\vec{x}^2}{r^2}
\end{equation}
where there are $d_s$ spatial dimensions $\vec x$, and we have set the curvature scale to one for convenience. The spacetime has an isometry under $t \to \lambda^z t$, $x \to \lambda x$, $r \to \lambda r$ which realizes the anisotropic scaling symmetry of the dual field theory.  The holographic dictionary for this spacetime generally has a similar structure to that for AdS (which corresponds to $z=1$) \cite{Ross:2009ar,Ross:2011gu} (see also \cite{Baggio:2011cp,Mann:2011hg}).  One of the significant differences is that while AdS had a smooth extension beyond $r=0$, the geometry \eqref{Lifspace} has a singularity there, as noted in \cite{Kachru:2008yh} and emphasized in \cite{Copsey:2010ya,Horowitz:2011gh}. Scalar curvature invariants constructed from \eqref{Lifspace} are necessarily finite -- indeed, constant -- as a consequence of the Lifshitz symmetry, but there are divergent tidal forces as we approach $r = 0$ along geodesic congruences. The consequences of this singularity for observers in the spacetime were explored in \cite{Horowitz:2011gh}, who argued that observers near the singularity would experience large effects, although in a particular model, \cite{Bao:2012yt} found that including the effects of the matter background made the effects on a test string finite.

However, it is not so clear how this singularity is reflected in observables in the dual field theory. The correlation functions of local field theory operators are not sensitive to the singularity, as they can be obtained from analytic continuation of Euclidean correlators, and the Euclidean spacetime is not singular. In \cite{Lei:2013apa} it was conjectured that the singularity could be reflected in the structure of the infrared divergences in scattering amplitudes. Scattering amplitudes are an intrinsically Lorentzian observable, and it is well-known that in massless theories they have infrared divergences associated with the emission of soft collinear particles. The singularity in the spacetime in the geometry \eqref{Lifspace} is related to the dual field theory having more soft modes, as the anisotropic scaling symmetry implies a dispersion relation $\omega \sim k^z$. The IR divergences in scattering amplitudes therefore seems a suitable place to look for observable effects of this physics.

The aim of the present paper is to investigate this by calculating the scattering amplitudes following the pioneering work of \cite{Alday:2007hr} in the AdS case. In that work, the scattering amplitude was related to the calculation  of a minimal surface (following \cite{Gross:1989ge} in the flat space case). The appropriate minimal surface was obtained in \cite{Alday:2007hr} by working in a T-dual geometry where it is a minimal surface ending on light-like segments on the asymptotic boundary of the T-dual spacetime, whose geometry is again AdS. This gives the leading behaviour of the amplitude as
 \begin{equation} \label{amplitude}
\mathcal{A} \sim e^{iS}
\end{equation}
where $S$ is the action of a string wrapping the minimal surface determined by the boundary conditions; this represents a stationary point approximation to the amplitude. In the case of $\mathcal N=4$ SYM, the scattering amplitude can be related to a Wilson loop \cite{Drummond:2007aua,Brandhuber:2007yx,Drummond:2007cf}, and \eqref{amplitude} can then be understood in terms of the saddle-point calculation of the dual Wilson loop; the leading IR singularity is then related to the cusp anomalous dimension \cite{Korchemsky:1985xj}. We are not claiming that such an amplitude-Wilson loop duality extends to the Lifshitz field theories; we simply want to use \eqref{amplitude} as a convenient trick to evaluate the leading behaviour of the amplitude, working in a T-dual frame because it's easier to find the minimal surface there, in the spirit of the discussion in  \cite{Alday:2007hr}. This can perhaps be made more rigorous in the context of the string embedding of $z=2$ Lifshitz spacetime  in \cite{Balasubramanian:2010uk,Donos:2010tu}, or in the construction of \cite{Gregory:2010gx}, but we will leave this as a problem for the future.

In the AdS case, the external states in the scattering amplitude have a dispersion relation $\omega = \pm k$ determined by the conformal invariance, and the amplitude is related to a closed polygonal Wilson loop made up of light-like segments whose lengths are related to the momenta of the external particle states. In \cite{Alday:2007hr}, the expectation value of the Wilson loop related to the four-point amplitude was shown to be related by conformal invariance to a simple case with two light like segments meeting in a cusp, originally analysed in  \cite{Kruczenski:2002fb}. In AdS, the minimal surface corresponding to this cusp is simply
 \begin{equation}\label{AdSsolution}
r^2=2(t^2-x^2),
\end{equation}
which satisfies the boundary conditions $r=0$ at $t = \pm x$.

We will consider the analogue of this cusp in the Lifshitz case. Since we have less symmetry than in AdS, this is no longer related to the Wilson loop with four segments, and finding the appropriate minimal surface in the bulk for a full scattering amplitude is much more difficult. But considering this cusp will suffice to enable us to control the leading IR singularity in the amplitude.

In section \ref{dual}, we will set up the calculation in Lifshitz. In the Lifshitz case, the anisotropic scaling symmetry determines the dispersion relation to be $\omega = \pm \alpha k^z$, where $\alpha$ is  an undetermined parameter which would be fixed by the microscopic details of the field theory. Since we can't control these details of the field theory, We will look for minimal surfaces satisfying $r=0$ at $t = \pm \alpha x^z$, treating $\alpha$ as a free parameter. The lines $t = \pm \alpha x^z$ are timelike in the boundary at $r \to 0$ for any $\alpha$ because of the non-relativistic causal structure in the boundary of \eqref{Lifspace}. Therefore in section \ref{timel} we will give a brief discussion of null and timelike cusps in the AdS case, to fix expectations for the behaviour of our results as $z \to 1$.

Then in section \ref{min} we find the minimal surfaces satisfying these boundary conditions, and determine the leading IR divergences in the amplitudes. We will find that these minimal surfaces have a peculiar ``mushroom'' shape, where the surface initially bends away to larger $x$ before turning around. The leading divergence is controlled by the near-boundary behaviour of the surface. We find this divergence is stronger than in the corresponding timelike cusps for $z=1$, with a universal ($z$-indepenent) dependence on the cutoff with a coefficient which vanishes as we take the limit. This result is reminiscent of the behaviour of the bulk singularity, where curvature diverges as $1/\tau^2$ along the worldline of geodesics which approach the singularity (where $\tau$ is the proper time), with a coefficient which vanishes as $z \to 1$ \cite{Copsey:2010ya}. It should be understood in the field theory as due to the presence of the higher density of soft modes implied by the modified dispersion relation $\omega \sim k^z$.

\section{Lifshitz amplitudes}
\label{dual}

We are interested in considering a scattering amplitude in the Lifshitz background \eqref{Lifspace}. This involves insertion of on-shell particles in the boundary field theory at $t = \pm \infty$. In the bulk, it is determined by a string world sheet located near the singularity $r = \infty$ in \eqref{Lifspace}; it is thus infrared divergent. This divergence can be cut off as in \cite{Alday:2007hr}, by considering a brane at some fixed $r=r_0$ (taking $r_0 \to \infty$ at the end of the calculation). The regulated amplitude is given by
\begin{equation} \label{amp}
A_n\sim <\prod_{i=1}^{n}V_i(\mathbf{k}_i,\mathbf{x}(\sigma_i))> \sim \int DX e^{iS}e^{i\sum_{i=1}^{n}\mathbf{k}_i\cdot \mathbf{x}(\sigma_i)}
\end{equation}
where \begin{equation} \label{ngs}
S=\frac{1}{2\pi}\int d\sigma d\tau [-\frac{1}{r^{2z}}(\partial_\alpha t)^2+\frac{1}{r^2}(\partial_\alpha x)^2+\frac{(\partial_\alpha r)^2}{r^2}]
\end{equation}
is the Nambu-Goto action of the string worldsheet,\footnote{If we were to do a proper top-down construction this should be replaced by an appropriate superstring action, but we will neglect such details; at least in the simplest AdS context the problem reduces to finding the minimal surface which extremizes \eqref{ngs} as we will do here.}  and the string world sheet has a boundary on the regulating brane at $r=r_0$, with Dirichlet boundary conditions on $r$ and Neumann boundary conditions on the field theory directions. We approximate the scattering amplitude by a saddle point which extremizes \eqref{ngs} subject to these boundary conditions.

As in \cite{Alday:2007hr}, the minimal surface is more easily obtained by working in the T-dual coordinates, T-dualizing along the boundary directions $t, \vec{x}$. We view this as a trick to obtain the minimal surface we are interested in living in the original spacetime, so we will not carefully investigate this T-duality transformation. This has been studied extensively in the AdS case \cite{Ricci:2007eq,Beisert:2008iq}, and some of those results may admit extensions to the Lifshitz context, at least in the context of the supersymmetric realizations of $z=2$ Lifshitz in \cite{Balasubramanian:2010uk,Donos:2010tu}, but we will not investigate this further.

T-dualizing \eqref{Lifspace} along $t$, $\vec{x}$ to T-dual coordinates $t', \vec{x}'$ gives us back a Lifshitz spacetime in the coordinates $t', \vec{x'}, r' = 1/r$, but with a different dilaton field\footnote{This expression will formally have an imaginary part due to T-dualizing the time direction. However, this dilaton does not affect the evaluation of the saddle-point minimal surface.}
$ \phi = (z+d_s) \ln r $. 
The minimal surface we wanted to find  thus becomes an extremum of the Nambu-Goto action \eqref{ngs} in terms of the T-dual coordinates, with a boundary at $r' = 1/r_0$ with Dirichlet boundary conditions in the $t', \vec{x}'$ directions. The momentum of the external states becomes separation in the $t', \vec{x}'$ directions, so the boundary of the string worldsheet is fixed to lie on a closed polygon at $r'=1/r_0$ made up of segments with $\Delta t' = \alpha |\Delta \vec x'|^z$ for some $\alpha$. In the limit $r_0 \to \infty$, this is a polygon in the boundary $r'=0$ of the T-dual spacetime.

Our main aim is to find the minimal surface satisfying these boundary conditions. Actually, this is rather difficult for a non-trivial polygon, so we will consider just the corner between two such segments; that is, we take the boundary conditions for our minimal surface to be $r' = 0$ at $t' = \pm \alpha |\vec{x}'|^z$, for $t' >0$. Since two segments define a plane, we can orient our coordinates such that the separation is just along one of the spatial directions in the Lifshitz metric \eqref{Lifspace}, and the minimal surface will lie in a three-dimensional subspace of \eqref{Lifspace}.

We will henceforth drop the primes on the dual coordinates. Our interest is then in finding a minimal surface in the three-dimensional subspace
\begin{equation}
ds^2=-\frac{dt^2}{r^{2z}}+\frac{dr^2+dx^2}{r^2}
\end{equation}
satisfying the cusp boundary conditions $r = 0$ at $t = \pm \alpha x^z$ for $t >0$.

One might be tempted to parametrize the surface by $t, x$, but we will find that it is actually not a single-valued function of $x$: the surface moves initially to larger $x$ as $r$ increases, before returning to smaller $x$. Using the fact that the action \eqref{ngs} is invariant under $x \to -x$, the surface satisfying our boundary conditions will be symmetric under $x \to -x$, so we can restrict attention to the surface for $x >0$. We can then parametrize the surface for $x >0$  by $t,r$. Using the scaling symmetry, a more convenient choice of parametrization of this surface is in terms of $\sigma, f$ where
 \begin{equation} \label{param}
t=\sigma^z; \qquad x=\sigma u(f); \qquad r=\sigma f \qquad (t\ge 0)
\end{equation}
Our task is to determine the form of $u(f)$ which extremizes the Nambu-Goto action, subject to the boundary condition $u(0) = u_0$ for some arbitrary parameter $u_0>0$,  where $u_0^z = \alpha$, and $u(f_0)=0$ at some $f_0 >0$.  The Nambu-Goto action is
\begin{eqnarray} \label{action}
S&=&\frac{1}{2\pi \alpha'} \int dX_a dX_b\sqrt{\det(G_{\mu\nu}\partial_aX^{\mu}\partial_b X^\nu)} \\ \label{lifaction}
 &= & \frac{1}{2\pi \alpha'} \int \frac{d\sigma}{\sigma} \int \frac{df}{f^{z+1}}\sqrt{(z^2-f^{2z})u'^2+2uf^{2z-1}u'+ (z^2-u^2f^{2z-2})}. \nonumber
\end{eqnarray}
The stationary point equation is
\begin{eqnarray}\label{equ}
\nonumber
	f[ f^2(f^{2z} - z^2) + f^{2z} u^2 ] u'' + f^2(z+1) (z^2 - f^{2z}) u'^3 \\
	+ f^{2z+1} (3z+1) u u'^2 + [ f^2(f^{2z} (z-1) + z^2 (z+1)) - 2z f^{2z} u^2 ]u' \\
\nonumber
	- (z-1) f^{2z+1} u = 0 .
\end{eqnarray}
Note that the points at which
\begin{equation}\label{Delta f}
	\Delta(f) = f[ f^2(f^{2z} - z^2) + f^{2z} u^2 ] = 0
\end{equation}
are singular points of \eqref{equ}. The choice of parametrization \eqref{param} thus reduces the problem to an ODE. This is a complicated non-linear ODE, but it is straightforward to solve numerically when $\Delta$ does not change sign.

\section{Timelike and null cusps in AdS spacetime}
\label{timel}

Before discussing the solutions of \eqref{equ} in the Lifshitz case, it is useful to briefly return to AdS by setting $z=1$. This equation then simplifies, and analytic solutions can be found for $u_0=1$, corresponding to a null cusp in the boundary. There are in fact two analytic solutions satisfying the boundary conditions, $u = \sqrt{1 - f^2/2}$, which corresponds to the solution \eqref{AdSsolution}, which is spacelike in the bulk, and $u = \sqrt{1-f^2}$, which corresponds to a null surface in the bulk spacetime. For $u = \sqrt{1-f^2}$, the action \eqref{action} vanishes identically. For $u = \sqrt{1-f^2/2}$,
\begin{equation}
S = \frac{1}{2 \pi \alpha'} \int \frac{d\sigma}{\sigma} \int \frac{df}{f^2} \sqrt{ - \frac{f^2}{2(2-f^2)}} \approx  \frac{i}{4 \pi \alpha'} \int \frac{d\sigma}{\sigma} \int \frac{df}{f},
\end{equation}
where in the second step we have kept the part that gives the leading divergence near $f=0$. We want to introduce a cutoff $\Delta t = \Delta x = \epsilon$ to regulate this divergence. This corresponds to cutting off the $\sigma$ integral at $\sigma_{\min} = \epsilon$, and cutting off the range of $u$ at $u_{max} = 1 - \frac{\epsilon}{\sigma}$, corresponding to $f_{min}^2 = 4 \frac{\epsilon}{\sigma}$. Thus the leading divergence is
\begin{equation}
S \sim \frac{i}{8 \pi \alpha'} (\ln \epsilon)^2
 \end{equation}

In the Lifshitz case, the boundary conditions correspond to a timelike cusp in the boundary, so it will be useful to understand the minimal surfaces for timelike cusps in AdS to facilitate the comparison for the $z \to 1$ limit of our results. This corresponds to taking $u_0 < 1$. Here we cannot find analytic solutions; we will first consider a series expansion and then find full solutions numerically. There is a  series expansion near $f=0$ which is valid for $u_0 \neq 1$,
\begin{equation}\label{asympt f=0 AdS}
	u(f) = u_0 + u_3 f^3 + \sum_{i=5} u_i f^i,
\end{equation}
where $u_0$ and $u_3$ are free data and the first few subleading terms are
\begin{equation}
	u_5 = - \frac{3 u_3}{5(u_0^2 - 1)}, \qquad u_6 = - \frac{2 u_3^2 u_0}{(u_0^2 - 1)}.
\end{equation}
Note that the general series expansion has no $O(f^2)$ term, so it cannot match smoothly on to the $u = (1-f^2/2)^{1/2}$ solution at $u_0 = 1$.

\begin{figure}[h]
\begin{center}
\includegraphics[scale=0.45]{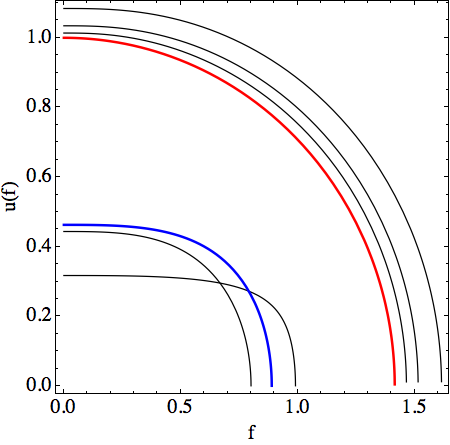}
\caption{The timelike ($u_0 <1$) and spacelike ($u_0 >1$) minimal surfaces in AdS. The black lines are extremal surfaces for $f_0 = 0.8$, $0.99$, $\sqrt{2} + 0.05$ $\sqrt{2} + 0.1$, $\sqrt{2} + 0.2$, the blue line corresponds to
$f_0 \approx 0.889 $ (which has maximum $u_0$ among our timelike surfaces) and the red line is the exact solution $u = (1-f^2/2)^{1/2}$. Note that $u_0$ is not a monotonic function of $f_0$ in the timelike case, its maximum being $u_0 \approx 0.463$.}
\label{fig:f0 free AdS}
\end{center}
\end{figure}

\begin{figure}[htb]
\center
\subfigure[][]{
\label{fig:L2 f0 free AdS}
\includegraphics[width=0.4\linewidth]{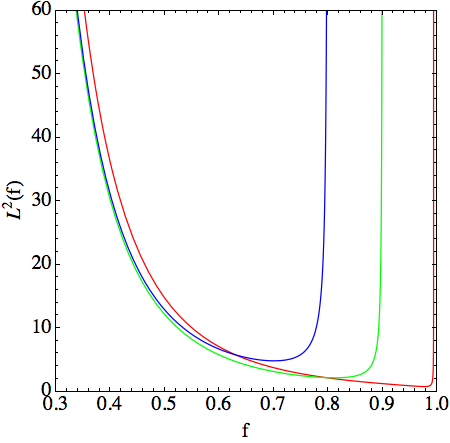}
}\qquad\qquad
\subfigure[][]{
\label{fig:L2 f0 free AdS space like}
\includegraphics[width=0.4\linewidth]{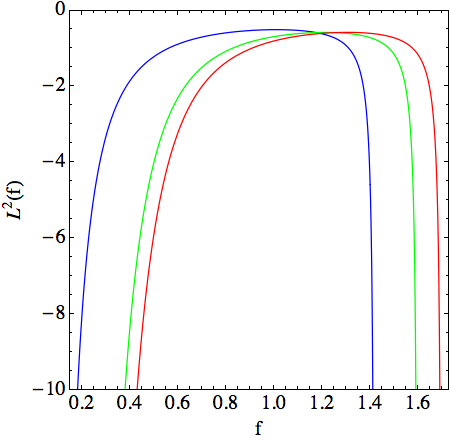}
}
\caption{(a): square of the on-shell Lagrangian for extremal surfaces in AdS with $f_0 = 0.8$ (blue), $f_0 =0.9$ (green), $f_0 =0.995$ (red). All surfaces are time-like ($L^2 > 0$) 
and they become almost null near the turning point as we approach $f_0 = 1$.
(b): square of the on-shell Lagrangian for extremal surfaces in AdS with $f_0 = 1.42$ (blue), $f_0 =1.6$ (green), $f_0 =1.7$ (red). All surfaces are space-like ($L^2 < 0$).}
\end{figure}

We construct solutions numerically by picking some value $f_0$ at which to take $u=0$, and integrating inwards towards $f=0$. Near $f_0$ we take an ansatz
\begin{equation}
	u(f)  = (f_0 - f)^{1/2} \sum_{i=0} b_i (f_0 - f)^{i},
\end{equation}
where $f_0$ is left undetermined by the equation of motion and the first coefficients are given by
\begin{equation}
	b_0 = \sqrt{f_0}, \qquad b_1 = \frac{5 - 4 f_0^2}{12 \sqrt{f_0}(f_0^2-1)}.
\end{equation}
For $f_0 = \sqrt{2}$, these coefficients agree with $u = (1-f^2/2)^{1/2}$. We find numerically that in the ranges $f_0 <1$, $f_0 > \sqrt{2}$, $\Delta$ has a definite sign, so we can construct the surfaces integrating inwards from $u=0$. We plot these surfaces in figure \ref{fig:f0 free AdS}. For $f_0 < 1$ they are time-like; they approach a null surface near the turning point as $f_0 \to 1$, see fig. \ref{fig:L2 f0 free AdS}. For  $f_0 > \sqrt{2}$ the minimal surfaces are space-like,  see fig. \ref{fig:L2 f0 free AdS space like}, and as $f_0 \to \sqrt{2}$ they seem to approach the analytic solution. We were unable to find solutions for  $1< f_0 < \sqrt{2}$ using either this simple radial integration or relaxation. It was remarked in \cite{Kruczenski:2002fb} that there are no minimal surfaces in AdS for $u_0$ approaching $1$ from below. We note that there is a range of $u_0$ values $\sim 0.463 < u_0 < 1$ where we find no minimal surfaces, in agreement with the claim of \cite{Kruczenski:2002fb}.
We do not have a physical understanding of the non-existence of surfaces with $1< f_0 < \sqrt{2}$. In principle, one could attempt to construct them by 
patching radial integrations from $u=0$ to the vicinity of the singular point $\Delta = 0$ and from the singular point towards the boundary. If they exist, we believe that 
they would change signature in the bulk, see figure \ref{fig:L2 f0 free AdS}. Extremal surfaces with non-definite signature have been found \cite{Gibbons:2004dz}, 
so we do not expect this feature to be an obstruction for their existence. We leave this investigation for future work.

\begin{figure}[htb]
\center
\subfigure[][]{
\label{fig:fit f0 free AdS}
\includegraphics[width=0.4\linewidth]{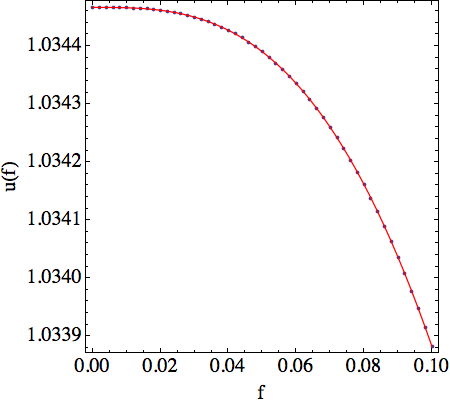}
}\qquad\qquad
\subfigure[][]{
\label{fig:fit f0 free AdS2}
\includegraphics[width=0.4\linewidth]{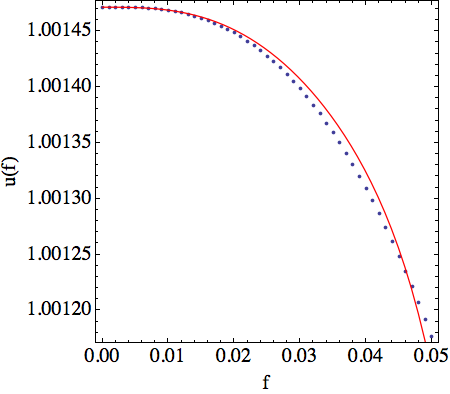}
}
\caption{Extremal surface for (a) $f_0 = \sqrt{2} +0.1$ and (b) $f_0 = \sqrt{2} +0.01$ in the region close to $f= 0$ and fit with the asymptotic expansion \eqref{asympt f=0 AdS}.
Since the integration from $f= f_0$ produces surfaces connected to $u = (1-f^2/2)^{1/2}$, the asymptotics \eqref{asympt f=0 AdS} do not fit well the data
for $f_0$ close to $\sqrt{2}$.}
\end{figure}

The leading divergence in the action comes from the behaviour near the boundary $f=0$. For $u_0 \neq 1$ the action simplifies to
\begin{equation}
S \approx \frac{1}{2\pi \alpha'} \int \frac{d\sigma}{\sigma} \int \frac{df}{f^2} \sqrt{ 1- u_0^2}
\end{equation}
If we cut off $u$ at $u =u_0 - \frac{\epsilon}{\sigma}$ as before, this now corresponds by \eqref{asympt f=0 AdS} to cutting off $f$ at $f_{min} = (\frac{\epsilon}{|u_3| \sigma})^{1/3}$, and\footnote{Note that we assume $u_3 <0$ to obtain a consistent form for $f_{min}$. This is consistent with our numerical solutions.}
\begin{equation}
S \sim \frac{\sqrt{1-u_0^2}}{\epsilon^{1/3}}.
\end{equation}
Note that the divergence for these timelike Wilson loops is stronger than in the null case. Note also that because of the $\sigma$ dependence in the cutoff for $f$, the integral over $\sigma$ is now finite; the leading divergence in the action for the timelike case is not concentrated in the vertex of the cusp, but comes from the limit of the range of $x$ at every $t$.

Thus, the leading divergence for the AdS surfaces with $u_0 \neq 1$ is stronger than in the case $u_0=1$ considered previously. For the surfaces with $u_0 > 1$, the action is imaginary, corresponding to an exponential suppression of the amplitude \eqref{amp}, and the coefficient of this stronger divergence will vanish as we approach the null case. For $u_0 <1$, the action is real because the surface is timelike.

Since our Lifshitz surfaces will always have timelike cusps on the boundary, we expect them to approach these timelike cusps in the limit as $z \to 1$.

\section{Minimal surfaces in Lifshitz}
\label{min}

We now turn to our main results, solving \eqref{equ} to find the minimal surfaces in Lifshitz giving a saddle-point approximation to the amplitudes. We will consider generic values of $z$, focusing on the range $1 < z < 2$. As usual, there will be some additional logarithmic terms arising for specific values such as $z=2$; we do not treat these but it should be straightforward to extend our analysis to these cases.

As in AdS, we can first consider an asymptotic expansion near the boundary $f=0$. In this case we find
\begin{equation}\label{solu1}
u(f)=u_0+\frac{(z-1)u_0}{2z^3(2-z)}f^{2z}+bf^{z+2}+...
\end{equation}
As $z \to 1$, the coefficient of the leading non-trivial term in the series vanishes, and we recover the expansion \eqref{asympt f=0 AdS} in the AdS case. Actually, the limit as $z \to 1$ of the asymptotic series expansion is somewhat subtle, as there are terms in the expansion with powers which coincide in the limit. We discuss this limit for the full series expansion in more detail in appendix \ref{series}.

From \eqref{solu1}  we see a remarkable feature of the Lifshitz minimal surfaces; the value of $u$ (and hence $x$ at fixed $t$) is initially increasing for any choice of the free parameter $b$. Thus, any solution consistent with this asymptotic series solution will initially move to increasing $u$ as we move into the interior of the spacetime, even though our boundary conditions imply the surface must reach $u=0$ at some finite $f$. Minimal surfaces satisfying these boundary conditions will thus have a ``mushroom'' shape. This is indeed what we find in our numerical analysis.

As in the AdS case, numerical solutions are found by starting from the point $f_0$ at which $u=0$ and integrating in. We find that the equation can be satisfied perturbatively near this point with an expansion of the form
\begin{equation}
	u(f) = (f_0 - f)^{1/2} \sum_{i= 0} b_i (f_0 - f)^{i},
\end{equation}
which reproduces the expected behaviour that $f'(u) = 0$ at $u=0$.  At the first non-trivial order in $(f_0 - f)$, the equation of motion implies
\begin{equation}
	b_0 [ 2 f_0 - (z+1) b_0^2 ](f_0^{2z} - z^2) = 0.
\end{equation}
This leads to three different branches of solutions characterized by
\begin{enumerate}
\item $ 2 f_0 - (z+1) b_0^2 = 0$,
\item $f_0^{2z} - z^2 = 0 $,
\item or $b_0 = 0$.
\end{enumerate}

\begin{figure}[htb]
\center
\subfigure[][]{
\label{fig:f0 free}
\includegraphics[width=0.4\linewidth]{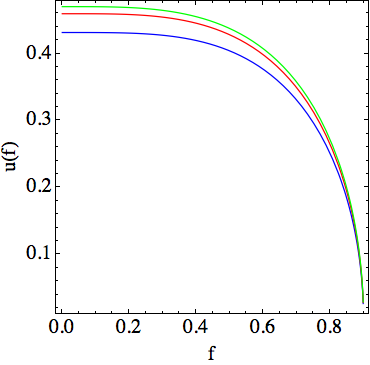}
}\qquad\qquad
\subfigure[][]{
\label{fig:mushroom z=9/8}
\includegraphics[width=0.46\linewidth]{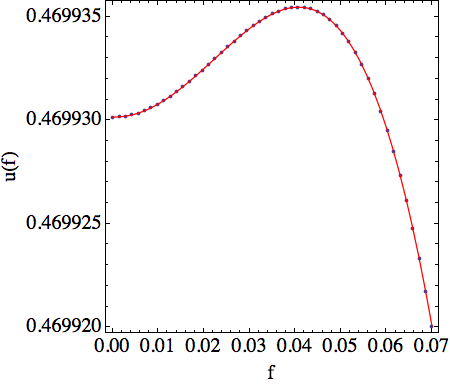}
}
\caption{(a): Extremal surfaces for $f_0 = 0.9$. The values of $z$ are (from top to bottom) $z=9/8, 5/4,3/2$.
(b): Extremal surface for $f_0 = 0.9$ and $z= 9/8$ in the region close to $f= 0$. The points are data obtained by numerical integration from $f = 0.9$,
while the solid line is the fit with the asymptotic expansion \eqref{solu1}.}
\end{figure}

We were able to find numerical solutions satisfying our boundary conditions only for the first case, in the range $f_0 < z^{1/z}$.  The equation of motion can then be easily integrated towards $f = 0$. Zooming in near the
$f= 0$ region, we note that $u'(f)$ changes sign, giving rise to surfaces with a ``mushroom" shape, see figures \ref{fig:f0 free}, \ref{fig:mushroom z=9/8}, as expected from the asymptotics. The behaviour near the boundary is consistent with the asymptotic expansion \eqref{solu1}. In the limit $z \to 1$, these solutions approach the timelike surfaces of section \ref{timel} with $f_0 <1$.

\begin{figure}[h]
\begin{center}
\includegraphics[scale=0.45]{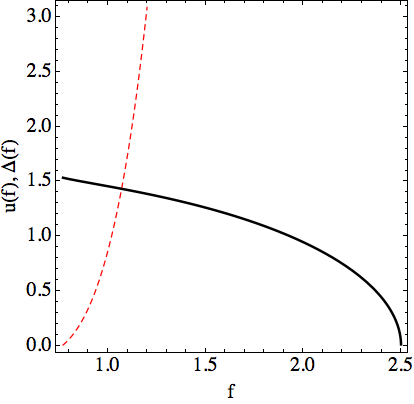}
\caption{For $f_0 = 2.5$ and $z= 3/2$, we plot part of the extremal surface (solid line) and $\Delta(f)$ in \eqref{Delta f} (dashed line). The vanishing of $\Delta$ prevents us from
continuing with the shooting from $f=f_0$.}
\label{fig:hit crit point}
\end{center}
\end{figure}

On the other hand, for $f_0 > z^{1/z}$ the integration encounters a critical point $\Delta(f) = 0$ before reaching $u=0$, see figure \ref{fig:hit crit point}. 
We also attempted to find solutions in this regime by a relaxation method, but this also fails to converge.
As mentioned in section \ref{timel}, one could attempt to construct these solutions by patching two shooting procedures.

The divergence of the action is determined by the near boundary expansion (\ref{solu1}). We are primarily considering the case $z<2$, where the second $2z$ term dominates. If we considered instead $z>2$, the third $z+2$ term would dominate the near-boundary expansion. In either case, $u' \approx 0$ near $f=0$, so the leading near-boundary contribution to the action is simply
\begin{equation}
S \approx \frac{z}{2\pi \alpha'} \int \frac{d\sigma}{\sigma} \int \frac{df}{f^{z+1}}.
\end{equation}
We want to impose a cutoff $\epsilon$, such that $\Delta x = \epsilon$, $\Delta t = \epsilon^z$. The $\sigma$ integral will thus have a lower bound $\epsilon$, while $u-u_0$ is bounded by $\frac{\epsilon}{\sigma}$, implying
\begin{equation}
f_{min} = \left[ \frac{2z^3 (2-z)}{(z-1) u_0} \frac{\epsilon}{\sigma} \right]^{1/2z}
\end{equation}
for $z<2$, and
\begin{equation}
f_{min} = \left[ \frac{1}{b} \frac{\epsilon}{\sigma} \right]^{1/(z+2)}
\end{equation}
for $z >2$. Thus, for $z<2$,
\begin{equation}
S \sim \frac{\sqrt{z-1}}{\sqrt{\epsilon}},
\end{equation}
while for $z >2$,
\begin{equation}
S \sim \frac{1}{\epsilon^{\frac{z}{z+2}}}.
\end{equation}
As in the timelike AdS case, this divergence is coming from the integral over $x$ at all $t$, and there is no additional divergence from the corner contribution at small $\sigma$.

We note that the divergence here is stronger than for the timelike surfaces in AdS, but with a coefficient which goes to zero in the limit as $z \to 1$, which is consistent with these minimal surfaces reducing to the ones in AdS in this limit. The mushroom feature in the shape of the surface also goes away in this limit, as can be seen from the expansion \eqref{solu1}. As remarked in the introduction, this stronger divergence in the Lifshitz case can be attributed to the presence of the higher density of soft modes implied by the modified dispersion relation $\omega \sim k^z$. It is interesting that this produces a divergence with a power that is independent of $z$ for $z<2$; this is consistent with the behaviour of the curvature singularity in the bulk for geodesic probes.

\section*{Acknowledgements} We are grateful for helpful discussions with Paul Heslop, Martin Kruczenski, Juan Maldacena and Reza Doobary. The work of TA and SFR is supported in part by STFC.

\appendix

\section{Asymptotic expansion for Lifshitz minimal surfaces}
\label{series}

In section \ref{min} we presented the leading terms in the asymptotic series around $f=0$. Here we discuss the full series and its behaviour as $z \to 1$. The general form of solution expansion is conjectured to be
\begin{equation}  \label{genralsol}
u(f)=u_0+\sum_{i,j=1}^{\infty}A_{ij}f^{a_{ij}}+\sum_{m,n=1}^{\infty}B_{mn}f^{b_{mn}},
\end{equation}
where the possible powers appearing in the expansion are
\begin{equation}\label{an}
a_{ij}=(2z-2)i+2+2z(j-1)
\end{equation}
and
\begin{equation}\label{bm}
b_{mn}=(2z-2)m+4-z+2z(n-1),
\end{equation}
and the first few coefficients are
\begin{equation}
A_{11}=\frac{(z-1)u_0}{2z^3(2-z)}; \qquad B_{11}=b
\end{equation}
\begin{equation}
A_{21}=\frac{(z-1)u_0^3}{2z^4(2z-1)(3z-4)(z-2)}; \qquad B_{21}=-\frac{bu_0^2(2+z)}{6z^3}
\end{equation}
\begin{equation}
A_{31}=\frac{(z-1)(2z-3)u_0^5}{2z^6(z-2)(5z-6)(3z-4)(3z-2)}
\end{equation}
\begin{equation} \label{ak1}
A_{k1}=\frac{u_0^{2k-1}(z-1)\prod_{\alpha=3}^{k}[(2\alpha-4)z-(2\alpha-3)]}{2z^{2k}[kz-(k-1)]\prod_{\beta=1}^{k}[(2\beta-1)z-2\beta ]} \qquad (k\ge 3)
\end{equation}
\begin{equation}
A_{12}=\frac{(z-1)(2z-1)(3z-1)u_0}{8z^6(3z-2)(2-z)}
\end{equation}

We note that this series expansion is valid for general values of $z$; there are special rational values where some of the denominators in these expressions for the coefficients vanish. At these values, two of the powers in \eqref{genralsol}, which are in general distinct, are coinciding. There will thus be log terms in the series expansion for these values. We do not consider them further here.

Now consider the limit as $z \to 1$. This limit is clearly very special for the above series expansion, as all the terms in the summations over $i$  and $m$ will have the same power of $f$ in the limit. We are particularly interested in the leading term at $j=1$, which would give a leading $f^2$ behaviour for the  asymptotic series expansion as $z \to 1$. From \eqref{ak1}, we see that each of these terms vanishes individually as we take the limit, but there are infinitely many of them, so it is not clear what the behaviour of the sum is in this limit.

Comparison to the asymptotic series \eqref{asympt f=0 AdS} in the AdS case would lead us to expect that the coefficient of the $f^2$ term will vanish as we take the limit for our minimal surface solutions with $u_0 <1$, and that is consistent with our numerical results, but here we want to consider if there is some other way to take the limit  that could converge to the solution $u = \sqrt{1-f^2/2}$ at $z=1$, which does have a non-trivial $f^2$ part in its asymptotic expansion.

We therefore consider the limit of \eqref{genralsol} assuming $u_0 \to 1$ as $z \to 1$. Let us write
 \begin{equation}\label{u0}
u_0=1+q\epsilon+\mathcal{O}(\epsilon^2),
\end{equation}
and $z = 1 + \epsilon$. We want to calculate
\begin{eqnarray}
A_{1}=\sum_{k=1}^{\infty} A_{k1} &=& \frac{(z-1)u_0}{2z^3(2-z)} +\frac{(z-1)u_0^3}{2z^4(2z-1)(3z-4)(z-2)} \\
&&+ \sum_{k=3}^{\infty}\frac{u_0^{2k-1}(z-1)\prod_{\alpha=3}^{k}[(2\alpha-4)z-(2\alpha-3)]}{2z^{2k}[kz-(k-1)]\prod_{\beta=1}^{k}[(2\beta-1)z-2\beta ]} \\
&=& \frac{\epsilon u_0}{2(1+\epsilon)^3(1-\epsilon)}+ \frac{\epsilon u_0^3}{2(1+\epsilon)^4(1+2\epsilon)(1-3\epsilon)(1-\epsilon)} \\
&&+\sum_{k=3}^{\infty}\frac{\epsilon u_0^{2k-1}\prod_{\alpha=3}^{k}[1-(2\alpha-4)\epsilon]}{2(1+\epsilon)^{2k}(1+k\epsilon)\prod_{\beta=1}^k[1-(2\beta-1)\epsilon]}
\end{eqnarray}
One can find that
\begin{eqnarray}
A_1&=&\sum_{k=1}^{\infty} \frac{\epsilon}{2}(1-2\epsilon)u_0^{2k-1} \\
&=& \frac{\epsilon}{2} (1-2\epsilon) \frac{u_0}{1-u_0^2}=-\frac{1}{4q}+ \mathcal{O}(\epsilon),
\end{eqnarray}
where we used (\ref{u0}) in the last step. Thus, for $u_0 \to 1$, the sum of the series can give a non-zero answer. Note that for $u_0 \neq 1$ the sum is zero, consistent with the expansion \eqref{asympt f=0 AdS} in the AdS case for $u_0 \neq 1$. To obtain the $u = \sqrt{1-f^2/2}$ solution in the limit, we would need $q =1$.

Thus, if there were solutions with $u_0 \to 1$ from above in the limit, they could be smoothly connected to the usual AdS minimal surface for the lightlike Wilson loop. However, numerically we have only found solutions for minimal surfaces in Lifshitz with $u_0 <1$. In the $z \to 1$ limit these converge to the timelike AdS surfaces discussed in section \ref{timel}.

\end{document}